\documentclass[nofootinbib,10pt,twocolumn,preprintnumbers]{revtex4-1}
\usepackage{amsmath,amssymb,pgf,pgfarrows,pgfnodes,float,appendix, hyperref,slashed,breakurl,graphicx}
\definecolor{Color}{rgb}{0.28, 0.24, 0.55}
\definecolor{Orange}{rgb}{1,0.38,0.11}
\hypersetup{
    colorlinks = true,
    citecolor = Orange,
    linkcolor = Color,
    urlcolor  = Orange,
}
\definecolor{internationalorange}{rgb}{1.0, 0.31, 0.0}

\usepackage{graphicx}
\usepackage{subfigure}
\usepackage[margin=0.9in]{geometry}

\usepackage{enumerate}

\usepackage{amsmath}

\newcommand{\SU}{\text{SU}}

\usepackage{color, colortbl}
\definecolor{Gray}{gray}{0.8}
\definecolor{GrayLight}{gray}{0.4}
\definecolor{Darkgreen}{RGB}{30,120,30}
\definecolor{granate}{rgb}{0.8039,0.2,0.2}
\newcommand{\beq}{\begin{equation}}
\newcommand{\eeq}{\end{equation}}
\newcommand{\bea}{\begin{eqnarray}}
\newcommand{\eea}{\end{eqnarray}}

\usepackage{eso-pic}

\usepackage{xparse}

\usepackage{tikz}
\usetikzlibrary{"arrows", "automata", "backgrounds", "calendar", "chains", "matrix", "mindmap", "patterns", "petri", "shadows", "shapes.geometric", "shapes.misc", "spy", "trees"}
\usetikzlibrary{arrows,shapes}
\usetikzlibrary{trees}
\usetikzlibrary{matrix,arrows} 				% For commutative diagram
\usetikzlibrary{positioning}				% For "above of=" commands
\usetikzlibrary{calc,through}				% For coordinates
\usetikzlibrary{decorations.pathreplacing}  % For curly braces
\usepackage{pgffor}							% For repeating patterns

\usetikzlibrary{decorations.pathmorphing}	% For Feynman Diagrams
\usetikzlibrary{decorations.markings}
\tikzset{
    vector/.style={decorate, decoration={snake}, draw},
	provector/.style={decorate, decoration={snake,amplitude=2.5pt}, draw},
	antivector/.style={decorate, decoration={snake,amplitude=-2.5pt}, draw},
    fermion/.style={draw=black, postaction={decorate},
        decoration={markings,mark=at position .55 with {\arrow[draw=black]{>}}}},
    fermionr/.style={draw=black, postaction={decorate},
    decoration={markings,mark=at position .55 with {\arrow[draw=black]{<}}}},
    fermioncyan/.style={draw=black, postaction={decorate},
        decoration={markings,mark=at position .55 with {\arrow[draw=cyan]{<}}}},
    fermiondif/.style={draw=black, postaction={decorate},
        decoration={markings,mark=at position .7 with {\arrow[draw=black]{>}}}},
            fermiondif2/.style={draw=black, postaction={decorate},
        decoration={markings,mark=at position .7 with {\arrow[draw=black]{<}}}},
    fermionend/.style={draw=black, postaction={decorate},
        decoration={markings,mark=at position 1 with {\arrow[draw=black]{>}}}},
    fermionuchannel2/.style={draw=black, postaction={decorate},
        decoration={markings,mark=at position .4 with {\arrow[draw=black]{>}}}},
    scalardif/.style={dashed,draw=black, postaction={decorate},
        decoration={markings,mark=at position .7 with {\arrow[draw=black]{>}}}},
    scalarend/.style={dashed,draw=black, postaction={decorate},
        decoration={markings,mark=at position 1 with {\arrow[draw=black]{>}}}},
    fermionbar/.style={draw=black, postaction={decorate},
        decoration={markings,mark=at position .55 with {\arrow[draw=black]{<}}}},
    fermionnoarrow/.style={draw=black},
    gluon/.style={decorate, draw=black,
        decoration={coil,amplitude=4pt, segment length=5pt}},
    scalar/.style={dashed,draw=black, postaction={decorate},
        decoration={markings,mark=at position .55 with {\arrow[draw=black]{>}}}},
    scalarcyan/.style={dashed,draw=black, postaction={decorate},
        decoration={markings,mark=at position .55 with {\arrow[draw=cyan]{>}}}},
    scalaruchannel1/.style={dashed,draw=black, postaction={decorate},
        decoration={markings,mark=at position .7 with {\arrow[draw=black]{>}}}},
                  scalaruchannel2/.style={dashed,draw=black, postaction={decorate},
        decoration={markings,mark=at position .4 with {\arrow[draw=black]{>}}}},
    scalarbar/.style={dashed,draw=black, postaction={decorate},
        decoration={markings,mark=at position .55 with {\arrow[draw=black]{<}}}},
    scalarnoarrow/.style={dashed,draw=black},
    electron/.style={draw=black, postaction={decorate},
        decoration={markings,mark=at position .55 with {\arrow[draw=black]{>}}}},
	bigvector/.style={decorate, decoration={snake,amplitude=4pt}, draw},
}

\NewDocumentCommand\semiloop{O{black}mmmO{}O{above}}
{%
\draw[#1] let \p1 = ($(#3)-(#2)$) in (#3) arc (#4:({#4+180}):({0.5*veclen(\x1,\y1)})node[midway, #6] {#5};)
}

\tikzstyle{block} = [draw, rectangle, 
    minimum height=3em, minimum width=6em]

\usepackage{enumitem}
\usepackage{tikz}
\usetikzlibrary{fit}
\tikzset{%
  highlight/.style={rectangle,rounded corners,color=granate,draw,text opacity =1,
    fill opacity=0.5,thick,inner sep=0pt}
}

\usepackage{xparse}
\NewDocumentCommand\loopv{O{black}mmmO{}O{above}}
{%
\draw[#1] let \p1 = ($(#3)-(#2)$) in (#3) arc (#4:({#4+360}):({0.5*veclen(\x1,\y1)})node[midway, #6] {#5};)
}

\usepackage{placeins}

\tikzset{
    cross/.pic = {
    \draw[rotate = 45] (-#1,0) -- (#1,0);
    \draw[rotate = 45] (0,-#1) -- (0, #1);
    }
}

\tikzset{
    square/.style={%
        draw=none,
        circle,
        append after command={%
            \pgfextra \draw[#1] (\tikzlastnode.north-|\tikzlastnode.west) rectangle 
                (\tikzlastnode.south-|\tikzlastnode.east);\endpgfextra}
    },
    square/.default=black
}

\tikzstyle{block} = [draw, rectangle, 
    minimum height=3em, minimum width=6em]

\usepackage{xparse}

\begin{document}

\preprint{CALT-TH/2023-025}

\title{\Large {\bf{Finite Naturalness and Quark-Lepton Unification}}}
\author{Pavel Fileviez P\'erez$^{1}$, Clara Murgui$^{2}$, Samuel Patrone$^{2}$, Adriano Testa$^{2}$, Mark B. Wise$^{2}$}
\affiliation{$^{1}$Physics Department and Center for Education and Research in Cosmology and Astrophysics (CERCA), 
Case Western Reserve University, Cleveland, OH 44106, USA \\
$^{2}$Walter Burke Institute for Theoretical Physics, California Institute of Technology, Pasadena, CA 91125}

\begin{abstract}We study the implications of finite naturalness in Pati-Salam models where $\SU(3)_C$ is embedded in $\SU(4)$. For the minimal realization at low-scale of quark-lepton unification, which employs the inverse seesaw for neutrino masses, we find that radiative corrections to the Higgs boson mass are at least $\delta m_h^2 / m_h^2 \sim {\cal O}(10^4)$. The one-loop contributions to the Higgs mass are suppressed by four powers of the hypercharge gauge coupling. 
We find that for the vector leptoquarks the naively leading part of the two-loop corrections cancel. We assume the Dirac Yukawa couplings for neutrinos are equal to the up-type quark Yukawa couplings as predicted in the minimal theory for quark-lepton unification. Despite these findings, the two-loop corrections still dominate the finite naturalness bound. We mention a way to relax the lower bound on the vector leptoquark mass and have $\delta m_h^2 / m_h^2 \sim {\cal O}(10^2)$. 
\end{abstract}

\maketitle
%%%%%%%%%%%%%%%%%%%%%%%%%
\section{Introduction}
%%%%%%%%%%%%%%%%%%%%%%%%%
The hierarchy problem (for some recent reviews see~\cite{Giudice:2013yca,Craig:2022uua}) has motivated many proposed extensions of the standard model (SM), including technicolor and low energy supersymmetry. This problem arises from the quadratic dependence of the Higgs mass \mbox{parameter} on the momentum cut-off. The absence of experimental evidence for new particles at the Large Hadron Collider (LHC) has cast doubt on the relevance of this issue. The original description of the hierarchy problem involved integrating out momentum shells~\cite{Wilson:1983xri}. Although this is a compelling \mbox{physical} picture, the quadratic divergences do not appear in all regulator approaches, for example in dimensional regularization. 

In extensions of the SM that contain particles with masses (or fields with expectation values) much greater than the weak scale there can be a  quadratic dependence of the Higgs mass on these quantities that is independent of the regularization and subtraction scheme. Demanding that these are not too large compared to the Higgs mass is the principle of finite naturalness~\cite{Farina:2013mla}. Since the hierarchy problem is not a \mbox{mathematical} inconsistency, it might be a red herring associated with how we approach the theory. Nonetheless, in this paper we take finite naturalness as a serious constraint on extensions of the SM. 

An appealing extension of the SM is the $\SU(5)$ grand unified theory~\cite{Georgi:1974sy}.
Here, the strongest finite \mbox{naturalness} constraints occur at tree level (the doublet-triplet splitting problem). There are also problematic radiative corrections to the Higgs mass from virtual super heavy gauge bosons. Since the gauge couplings are known, these radiative corrections give $\delta m_h^2 / m_h^2 \sim {\cal O}( 10^{24})$, which is clearly \mbox{unacceptable} if finite naturalness is to be taken seriously.

Quark-lepton unification, introduced by Pati and Salam~\cite{Pati:1974yy,Pati:1973rp}, is also a very attractive extension of the SM. They proposed that the standard model $\SU(3)_C$ is embedded in $\SU(4)$ where the leptons are interpreted as the fourth color. 
In this theory, baryon number is conserved and therefore the vector leptoquarks do not have to be super heavy to \mbox{satisfy} the strong constraints on baryon number violating processes. However, the vector leptoquarks do give rise to new flavor-violating processes that are very constrained by experimental data. For \mbox{example}, in the absence of large mixings, $K_L \to \mu^\pm e^\mp$ \mbox{constrains} the mass of the vector leptoquarks to be above 1000 TeV~\cite{Valencia:1994cj}. Using the freedom in the unknown mixings between quarks and leptons this bound can be reduced to around $100$ TeV~\cite{Smirnov:2018ske,Gedeonova:2022iac}. 
\\
\\
In this article, we study the implications of finite \mbox{naturalness} on the minimal theory of quark-lepton unification based on the gauge group $\SU(4) \otimes \SU(2)_L \otimes {\rm U}(1)_R$~\cite{Smirnov:1995jq} (see also Ref.~\cite{FileviezPerez:2013zmv}). To achieve a low-scale breaking of $\SU(4)$, we use the inverse seesaw mechanism for neutrino masses~\cite{FileviezPerez:2013zmv}. For $100$ TeV vector leptoquarks, we find that %1-loop contributions to the Higgs mass, dominated by the neutral heavy gauge boson, lead to 
    $\delta m_h^2 / m_h^2 \sim {\cal O}(10^4)$, which is problematic from the perspective of finite naturalness. %We leave it up to the reader to decide to what extent this is problematic. 
In the two-loop corrections to the Higgs mass involving the vector leptoquark we show that the naively leading contributions cancel. However, we find that the remaining parts still dominate the overall radiative contribution to the Higgs mass.
An analogous effect happens in the two-loop contributions of the new neutral gauge boson, in which the naively leading corrections cancel. Finally, we discuss a way to relax the bounds on the vector leptoquark mass and have $\delta m_h^2 / m_h^2 \sim {\cal O}(10^2)$.
%%%%%%%%%%%%%%%%%%%%%%%%%%%%%%%%%%%%%%%%%%%%%%%%%%%%%
\section{Minimal Theory For Quark-Lepton Unification}
\label{sec:Minimal_Theory_QL}
%%%%%%%%%%%%%%%%%%%%%%%%%%%%%%%%%%%%%%%%%%%%%%%%%%%%%
We consider the simplest Pati-Salam type extension of the SM, based on the gauge group~\cite{Smirnov:1995jq,FileviezPerez:2013zmv}
\begin{equation}
  {\cal G}_{\rm PS} \equiv  \SU(4) \otimes \SU(2)_L \otimes {\rm U}(1)_R\,. 
\end{equation}
The quarks and leptons are unified in the following representations:
\begin{eqnarray}
   F_{\rm QL} &=& \begin{pmatrix}  u_L  &  \nu_L  \\ d_L & e_L   \end{pmatrix}   \sim (4,2,0)_{\rm PS}\,, \nonumber \\
   F_{u} &=& \begin{pmatrix} u_R  & \nu_R \end{pmatrix}  \, \,  \sim ( 4,1,1/2)_{\rm PS}\,, \\
   F_d &=& \begin{pmatrix} d_R & e_R \end{pmatrix} \, \,  \sim ( 4,1,-1/2)_{\rm PS}\,. \nonumber 
\end{eqnarray}
Each of these comes in three copies.
These representations contain the SM fermions plus (three) right-handed neutrinos. Since ${\cal G}_{\rm PS}$ is a product of three gauge groups, this theory has three gauge couplings, $g_4$, $g_2$, and $g_R$, which are determined by the SM gauge couplings $g_3$, $g_2$, and $g_Y$.

The gauge group ${\cal G}_{\rm PS}$ is connected to the SM gauge group through one breaking step, where an electrically neutral component of a non-trivial representation of $\SU(4)$ gets a vacuum expectation value (vev). This representation should be a singlet under $\SU(2)_L$, which is the same as the $\SU(2)_L$ in the SM. 
We will focus on the simplest scalar representation able to trigger such a breaking, 
 $\chi = (\chi^\alpha, \chi^4) \sim (4,1,1/2)_{\rm PS}$, and take $\langle \chi^A \rangle = \delta^{A4} v_\chi/\sqrt{2}$. Here $\alpha$ is a color index ($\alpha=1,2,3$) and $A$ is a $\SU(4)$ index ($A = 1, 2,3, 4$). The standard model hypercharge is given by
\begin{equation}
    Y = R + \frac{\sqrt{6}}{3} T_{15}\,,
\end{equation}
where $R$ is the ${\rm U}(1)_R$ charge, and $T_{15}$ is the $\SU(4)$ generator
\begin{equation}\label{eq:T15}
   T_{15}= \frac{1}{2\sqrt{6}} \text{diag}(1,1,1,-3)\,.
\end{equation}
%%%%%%%%%%%%%%%%%%%%%%%%%%%%%%%%%%%%%%%%%%%%%%%%
\subsection{The quark-lepton unification angle}
%%%%%%%%%%%%%%%%%%%%%%%%%%%%%%%%%%%%%%%%%%%%%%%
In this section, we summarize some aspects of \mbox{minimal} quark-lepton unification.
Once $\chi$ gets a vev, the new massive vectors associated with the broken \mbox{generators} of $\SU(4)$ acquire mass through the \mbox{covariant} derivative of $\chi$ 
\begin{eqnarray}
D^\mu \chi &=& \partial^\mu \chi + i g_4 T_a V_a^\mu \chi + i  \frac{g_R}{2} B_R^\mu \chi\,,
\end{eqnarray}
with $T_a$ being the generators of $\text{SU}(4)$, normalized in the standard fashion, i.e. $\text{Tr}(T_a T_b) = \frac{1}{2} \delta_{ab}$. 
The mass of the vector leptoquarks, $X_\mu \sim (3,1,2/3)_{\rm SM}$, is given by
\begin{equation}
 m_{X}^2   = \frac{1}{4} g_4^2 v_\chi^2\,.
\end{equation}
The new neutral massive gauge boson, $Z'_\mu$, is a linear combination of the $V_{15\mu}$ gauge boson associated to the broken $\SU(4)$ generator $T_{15}$ (see Eq.~\eqref{eq:T15}) and the $\text{U}(1)_R$ gauge boson, $B_{R\mu}$. Its mass comes from the $\chi$ kinetic term,
\begin{equation}
\begin{split}
{\cal L}&\supset (D_\mu \chi)^\dagger (D^\mu \chi)  \\
 &\supset  \frac{v_\chi^2}{8} \begin{pmatrix}  V_{15\mu} & B_{R\mu}  \end{pmatrix} \begin{pmatrix}   \dfrac{3g_4^2}{2}  & - \dfrac{3g_R g_4 }{\sqrt{6}}  \\  -\dfrac{3 g_R g_4 }{\sqrt{6}} &g_R^2   \end{pmatrix} \begin{pmatrix}  V_{15}^\mu \\ B_R^\mu \end{pmatrix}\,.
\end{split}
\end{equation}
The rotation matrix that brings the gauge bosons to their mass eigenbasis is
\begin{equation}\label{eq:rotationS}
\begin{pmatrix} V_{15\mu}  \\ B_{R\mu} \end{pmatrix} = \begin{pmatrix} \cos \theta_S & \sin \theta_S \\ -\sin \theta_S & \cos \theta_S \end{pmatrix} \begin{pmatrix}   Z_\mu' \\ B_\mu \end{pmatrix}\,,
\end{equation}
where the quark-lepton unification angle $\theta_S$ is given by
\begin{equation}
\sin \theta_S =\frac{g_R}{\sqrt{g_R^2 + \tfrac{3}{2} g_4^2}}\,, \quad \cos \theta_S = \frac{ g_4}{\sqrt{\tfrac{2}{3}g_R^2 + g_4^2}}\,.
\end{equation}
The hypercharge gauge coupling is  $g_1 = g_R \cos \theta_S$, while the strong gauge coupling is $g_3 = g_4$. Thus, the quark-lepton unification angle can be written as a function of standard model gauge couplings and the weak angle
\begin{equation}
    \sin^2 \theta_S = \frac{2}{3}\frac{\alpha_{\rm em}}{\alpha_s} \frac{1}{\cos^2\theta_W}\,.
\end{equation}
The mass of the gauge boson $Z'$ is given by the trace of the mass matrix (the hypercharge gauge boson is massless in the electroweak symmetric phase),
\begin{equation}
m_{Z'}^2 = \frac{1}{4}\left (\frac{3}{2} g_4^2 + g_R^2\right )v_\chi^2\,.
\end{equation}
Therefore,
\begin{equation}\label{eq:MZMXratio}
    \frac{m_X^2}{m_{Z'}^2} = \frac{2}{3}\cos^2\theta_S < 1\,.
\end{equation}
%
%%%%%%%%%%%%%%%%%%%%%%%%%%%%%%%%%%%%%%%%%%%%%%%%%%%%
\subsection{One-loop corrections to the Higgs mass}
%%%%%%%%%%%%%%%%%%%%%%%%%%%%%%%%%%%%%%%%%%%%%%%%%%%%
The Higgs boson is a singlet under $\SU(4)$. However, it carries $R = 1/2$ charge, $H_1 \sim (1,2,1/2)_{\rm PS}$, and therefore couples to the $Z_\mu'$. From the lower bound on the vector leptoquark mass mentioned in the introduction, Eq.~\eqref{eq:MZMXratio} implies $m_{Z'} > 130$ TeV.

The standard model Higgs couples to the $Z'_\mu$ gauge boson through the Higgs covariant derivative\footnote{The models of Refs.~\cite{Smirnov:1995jq,FileviezPerez:2013zmv} contain two Higgs doublets that live in two different representations of the ${\cal G}_{\rm PS}$. However, the results in this section are not affected.} resulting in the following Feynman rules
\begin{eqnarray}\label{eq:FRZH}
\begin{gathered} 
\begin{tikzpicture}[line width=1.5 pt,node distance=1 cm and 1 cm]
\coordinate[label=left:$H_1(p)$](h01);
\coordinate[right = 0.5 cm of h01](v1);
\coordinate[above right = 0.75 cm of v1,label=right:$Z_\mu'$](Zp);
\coordinate[below right = 0.75 cm of v1,label=right:$H_1(k)$](h02);
\draw[fill=black](v1) circle (.05cm);
\draw[scalarnoarrow](h01)--(v1);
\draw[scalarnoarrow](v1)--(h02);
\draw[vector](v1)--(Zp);
\end{tikzpicture}
\end{gathered} &:&  i\frac{ g_R}{2} (p + k)_\mu \sin \theta_S\,,\\
\begin{gathered}
\begin{tikzpicture}[line width=1.5 pt,node distance=1 cm and 1 cm]
\coordinate[label=left:$H_1$](h01);
\coordinate[below right = 0.75 cm of h01](v1);
\coordinate[below left = 0.75 cm of v1,label=left:$H_1$](h02);
\coordinate[above right = 0.75 cm of v1,label=right:$Z_\mu'$](Zp1);
\coordinate[below right = 0.75 cm of v1, label=right:$Z'_\nu$](Zp2);
\draw[fill=black](v1) circle (.05cm);
\draw[scalarnoarrow](h01)--(v1);
\draw[scalarnoarrow](v1)--(h02);
\draw[vector](v1)--(Zp1);
\draw[vector](v1)--(Zp2);
\end{tikzpicture}
\end{gathered}  &:&  i  2!\frac{g_R^2}{4} \sin^2\theta_S g^{\mu \nu}\,.
\end{eqnarray}
Therefore, the $Z_\mu'$ contributes to the Higgs boson mass through the following one-loop diagrams
\begin{eqnarray}
 && \!\!\!\!\!\!\!\!\!\!\!\! \begin{gathered} 
\begin{tikzpicture}[line width=1.5 pt,node distance=1 cm and 1 cm]
\coordinate[label=left:$H_1$](H1);
\coordinate[right = 1.5 cm of H1](v1);
\coordinate[right= 1.5 cm of v1,label=right:$H_1$](H2);
\coordinate[above=0.4 cm of v1,label=$\, \, \,Z'$](Zp);
\draw[fill=black](v1) circle (.05cm);
\draw[decorate, decoration={snake}] 
        (v1) arc (-90:270:0.7);
\draw[scalar](H1)--(v1);
\draw[scalar](v1)--(H2);
\end{tikzpicture}
\end{gathered} \nonumber\\
i \left. \delta m_h^2\right|_1 &=& i  \frac{g_R^2\sin^2 \theta_S}{4} \, g_{\mu \nu}  \int \frac{d^4k}{(2\pi)^4} \frac{-ig^{\mu \nu}}{k^2-m_{Z'}^2}\nonumber\\
&=&-i g_R^2 \sin^2 \theta_S \frac{m_{Z'}^2}{(4\pi)^2} \log \left(\frac{m_{Z'}^2}{m_h^2}\right)\,,
\end{eqnarray}
and 
\begin{eqnarray}
&&  \!\!\!\!\!\!\!\!\!\!\!\! \begin{gathered} 
\begin{tikzpicture}[line width=1.5 pt,node distance=1 cm and 1 cm]
\coordinate[label=left:$H_1$](H1);
\coordinate[right = 0.75 cm of H1](v1);
\coordinate[right = 2 cm of v1](v2);
\coordinate[right = 1 cm of v1](aux);
\coordinate[above = 0.2 cm of aux,label=$Z'$](Zp);
\coordinate[right = 0.75 cm of v2,label=right:$H_1$](H2);
\draw[fill=black](v1) circle (.05cm);
\draw[fill=black](v2) circle (.05cm);
\draw[scalar](H1)--(v1);
\draw[scalar](v1)--(v2);
\draw[scalar](v2)--(H2);
\semiloop[vector]{v1}{v2}{0};
\end{tikzpicture}
\end{gathered} \nonumber\\
i\left. \delta m_{h}^2 \right|_2 &=& i^2 g_R^2 \frac{\sin^2 \theta_S}{4} \int \frac{d^4k}{(2\pi)^4}\frac{i(-i)k^2}{(k^2-m_{Z'}^2)k^2}\nonumber\\
&=&i \frac{g_R^2}{4} \sin^2 \theta_S \frac{m_{Z'}^2}{(4\pi)^2} \log \left(\frac{m_{Z'}^2}{m_h^2}\right),
\end{eqnarray}
where we used Feynman gauge and set the subtraction point equal to the Higgs mass. In this paper, we neglect scheme-dependent terms that do not contain the large logarithm.
Adding up both contributions, we find that the one-loop correction to the Higgs boson mass is given by
\begin{equation}\label{eq:mh21loop}
\begin{split}
 & \left| \frac{\delta m_h^2 }{m_h^2} \right|= \frac{3}{4} \frac{g_R^2 \sin^2 \theta_S}{16\pi^2} \frac{m_{Z'}^2}{m_h^2}   \log \left(\frac{m_{Z'}^2}{m_h^2}\right)\\
   & \quad \simeq 4 \times 10^2 \left(\frac{m_{Z'}}{130 \text{ TeV}}\right)^2 \left(\frac{\log(m_{Z'}^2/m_h^2)}{13.9}\right).
\end{split}
\end{equation}
Note that the above result is suppressed by four \mbox{powers} of the hypercharge coupling since
\begin{eqnarray}
    g_R^2 \sin^2\theta_S &=&   \frac{8\pi}{3} \frac{\alpha_{\rm em}^2}{\cos^4\theta_W \alpha_s}\left(1-\frac{2}{3}\frac{\alpha_{\rm em}}{\alpha_s \cos^2 \theta_W}\right)^{-1} \nonumber \\
    &\simeq& 6 \times 10^{-3}\,.
\end{eqnarray}
%
%%%%%%%%%%%%%%%%%%%%%%%%%%%%%%%%%%%%%%%%%%%%%%%%%%
\subsection{Two-loop corrections to the Higgs mass}
\label{sec:2loop}
%%%%%%%%%%%%%%%%%%%%%%%%%%%%%%%%%%%%%%%%%%%%%%%%%% 
The dominant two-loop corrections to the Higgs mass contain interactions of the vector leptoquarks and the $Z'$ gauge boson with the fermions. We will give an estimate of the size of these contributions by computing only the diagrams involving the vector leptoquarks. 

Because of the vertex of the $Z'$ gauge boson with the Higgs, there are extra two-loop diagrams contributing to the corrections involving the neutral gauge boson. We do not consider these contributions given that they are proportional to powers of the weak coupling rather than the strong coupling. 

In quark-lepton unification, the Higgs Yukawa couplings predict the following mass matrix relations
\begin{equation}\label{eq:massrel}
    M_d = M_e\,,
    \quad \text{and} \quad M_u=M_\nu^D\,.
\end{equation}
\\
The main corrections to the Higgs mass will involve the top quark and the neutrinos, and therefore we focus on the $M_u = M_\nu^D$ prediction. The other relation in Eq.~\eqref{eq:massrel} will be addressed in the next section. 

To correct the $M_u = M_\nu^D$ identity, we assume the neutrinos get mass through the inverse seesaw~\cite{Mohapatra:1986aw,Mohapatra:1986bd}, adding three left-handed singlets $N_L$, which allows for the following interactions~\cite{FileviezPerez:2013zmv}
\begin{equation}
   - {\cal L} \supset Y_1 \bar F_{QL} \widetilde H_1 F_u +  Y_5 \bar F_u \chi N_L + \frac{1}{2}\mu  N_L^T C  N_L + \text{h.c.}\,,
\end{equation}
where $\widetilde H_1 = i \sigma_2 H_1^*$. In the broken phase, the mass matrix in the basis $(\nu_L,(\nu_R)^c,N_L)$ is given by
\begin{eqnarray}
   \begin{pmatrix} 0 & M_\nu^D & 0 \\ (M_\nu^D)^T & 0 & M_\chi^D \\ 0 & (M_\chi^D)^T & \mu \end{pmatrix}\,,
\end{eqnarray}
where $M_\nu^D = Y_1 v / \sqrt{2}$, $M_\chi^D = Y_5 v_\chi / \sqrt{2}$, and $v= 246$ GeV is the electroweak vev. We assume the following hierarchy $\mu \ll M_\nu^D \ll M_\chi^D$ which ensures that the light neutrinos are mostly $\nu_L$. We work in the limit $\mu \to 0$ and $M_\nu^D \ll M_\chi^D$, where the heavy neutrino mass eigenstates are approximately Dirac and are obtained by diagonalizing $M_\chi^D$. 

In the mass eigenbasis, choosing the left-handed neutrinos to align with the left-handed up-type quarks, the relevant interactions and mass terms are 
\begin{align}
-{\cal L} &\supset  -\frac{g_4}{\sqrt{2}} X_\mu \left(\bar u_L \gamma^\mu \nu_L  + \bar u_R \gamma^\mu W_R \nu_R \right) \nonumber\\
& + \frac{\sqrt{2}}{v} h\left( \bar u_L \widehat M_u u_R  + \bar \nu_L \widehat M_u W_R \nu_R \right) \\
& + \bar \nu_R \widehat M_\chi^D N_L + \bar u_L \widehat M_u u_R +  {\rm h.c.}\,,\nonumber
\end{align}
where $\widehat M_u = U_R^\dagger M_u U_L = \text{diag}(m_u,m_c,m_t)$, and the diagonal Dirac neutrino mass matrix $\widehat{M_{\chi}^D} = V_R^\dagger M_\chi^D V_L$. In the above equation,  $W_R = U_R^\dagger V_R$.
\newline

Now, we can proceed with the discussion of the two-loop contributions of the $X$ leptoquarks to the Higgs mass. We will first take all the couplings equal to one and ignore combinatorial factors. We identify two diagram topologies involving the vector leptoquarks and the fermions that give radiative corrections to the Higgs mass. We call $I_A$ the two-loop integral corresponding to topology A, given by
\begin{widetext}
\begin{eqnarray}\label{eq:topologyI}
  \begin{gathered} 
\begin{tikzpicture}[line width=1.5 pt,node distance=1 cm and 1 cm]
\coordinate[](h1);
\coordinate[right = 0.75 cm of h1](v1);
\coordinate[right = 2 cm of v1](v2);
\coordinate[right = 0.75 cm of v2](h2);
\coordinate[right = 1 cm of v1](vaux);
\coordinate[above =0.1 cm of vaux](Zp);
\coordinate[above = 1 cm of vaux](nu);
\coordinate[below = 1 cm of vaux](t);
\coordinate[above right = 1 cm of vaux](v4);
\coordinate[above left = 1 cm of vaux](v3);
\semiloop[vector]{v4}{v3}{180};
\draw[fill=black](v1) circle (.05cm);
\draw[fill=black](v2) circle (.05cm);
\draw[fill=black](v3) circle (.05cm);
\draw[fill=black](v4) circle (.05cm);
\loopv[fermionr]{v1}{v2}{0};
\draw[scalarnoarrow](h1)--(v1);
\draw[scalarnoarrow](v2)--(h2);
\end{tikzpicture}
\end{gathered} \qquad i \, I_{\rm A} &=& (-) \int \frac{d^4 {k}}{(2\pi)^4}    \int \frac{d^4{q}}{(2\pi)^4} \frac{\text{Tr}\{ (i \slashed{k}) \gamma_\mu (-ig^{\mu \nu}) i(\slashed{k}-\slashed{q}) \gamma_\nu (i \slashed{k}) (i \slashed{k}) P_{L,R}\}}{ (k-q)^2 (q^2-M^2) k^6}\,, \nonumber\\
\end{eqnarray}
while the two-loop integral associated with topology B is given by
\begin{eqnarray}\label{eq:topologyII}
\qquad
\begin{gathered}
\begin{tikzpicture}[line width=1.5 pt,node distance=1 cm and 1 cm]
\coordinate[](h1);
\coordinate[right = 0.75 cm of h1](v1);
\coordinate[right = 2 cm of v1](v2);
\coordinate[right = 0.75 cm of v2](h2);
\coordinate[right = 1 cm of v1](vaux);
\coordinate[above = 1 cm of vaux](nu);
\coordinate[below = 1 cm of vaux](t);
\coordinate[above right = 0.9 cm of vaux](v4);
\coordinate[above left = 1.5 cm of vaux](v3);
\draw[vector](t)--(nu);
\draw[fill=black](v1) circle (.05cm);
\draw[fill=black](v2) circle (.05cm);
\draw[fill=black](t) circle (.05cm);
\draw[fill=black](nu) circle (.05cm);
\loopv[fermionr]{v1}{v2}{0};
\draw[scalarnoarrow](h1)--(v1);
\draw[scalarnoarrow](v2)--(h2);
\end{tikzpicture}
\end{gathered} \qquad \!\!
i \, I_{\rm B} &=& (-) \int \frac{d^4 {k}}{(2\pi)^4} \int \frac{d^4{q}}{(2\pi)^4} \frac{\text{Tr}\{ (i \slashed{k}) (i \slashed{k}) \gamma_\mu (-ig^{\mu \nu}) i(\slashed{k}-\slashed{q})i(\slashed{k}-\mathbf{\slashed{q}}) \gamma_\nu P_{L,R} \}}{ (k-q)^4 (q^2-M^2) k^4}\,. \nonumber\\
\end{eqnarray}
\end{widetext}
We work in the limit where the relevant mass is the gauge boson mass ($M$), and consider the fermions massless. Notice that we include a projector in the trace to account for the chirality characterizing the fermionic line. For the purposes of this paper we treat $\gamma_5$ as anticommuting with all other $\gamma$ matrices. We evaluate the above expressions in $d = 4 -\epsilon$ dimensions. To assess the finite naturalness of this model we need to compute the leading scheme-independent finite parts of the radiative corrections to the Higgs mass.

By dimensional analysis, we expect all these diagrams to be proportional to 
\begin{equation*}
M^{2-2\epsilon} \underset{\epsilon \to 0}{\simeq} M^2 \left(1 - \epsilon \log M^2 + \frac{\epsilon^2}{2}\log^2 M^2\right)\,.
\end{equation*}
The $1/\epsilon^2$ contribution from each diagram multiplying the above generates a divergent term of the form $\left(1/\epsilon\right) M^2\log(M^2)$. Should such term survive in the sum of all the diagrams it would require non-analytic counterterms. Hence we expect no residual $1/\epsilon^2$ dependence in the overall amplitude. As a consequence, the leading finite part will depend linearly rather than quadratically on $\log M^2$. We explicitly checked that this cancellation occurs.

First, consider the diagrams 
\begin{equation}\label{diag:potato}
 \begin{gathered}
\begin{tikzpicture}[line width=1.5 pt,node distance=1 cm and 1 cm]
\coordinate[label=left:$h$](h1);
\coordinate[right = 0.25 cm of h1](v1);
\coordinate[right = 2 cm of v1](v2);
\coordinate[above left = 0.3 cm of v1,label=above:$t_L$](tLL);
\coordinate[above right = 0.3 cm of v2,label=above:$t_L$](tLR);
\coordinate[right = 0.3 cm of v2,label=right:$h \text{, }$](h2);
\coordinate[right = 1 cm of v1](vaux);
\coordinate[above =0.1 cm of vaux,label=$X$](Zp);
\coordinate[above = 1 cm of vaux,label=$\nu_{L}^3$](nu);
\coordinate[below = 1 cm of vaux,label=$t_R$](t);
\coordinate[above right = 1 cm of vaux](v4);
\coordinate[above left = 1 cm of vaux](v3);
\semiloop[vector]{v4}{v3}{180};
\draw[fill=black](v1) circle (.05cm);
\draw[fill=black](v2) circle (.05cm);
\draw[fill=black](v3) circle (.05cm);
\draw[fill=black](v4) circle (.05cm);
\loopv[fermionr]{v1}{v2}{0};
\draw[scalarnoarrow](h1)--(v1);
\draw[scalarnoarrow](v2)--(h2);
\end{tikzpicture}
\end{gathered} \!\!
 \begin{gathered}
\begin{tikzpicture}[line width=1.5 pt,node distance=1 cm and 1 cm]
\coordinate[label=left:$h$](h1);
\coordinate[right = 0.25 cm of h1](v1);
\coordinate[right = 2 cm of v1](v2);
\coordinate[right = 0.25 cm of v2,label=right:$h$](h2);
\coordinate[right = 1 cm of v1](vaux);
\coordinate[above left = 0.3 cm of v1,label=above:$t_R$](tRL);
\coordinate[above right = 0.3 cm of v2,label=above:$t_R$](tRR);
\coordinate[above =0.1 cm of vaux,label=$X$](Zp);
\coordinate[above = 1 cm of vaux,label=$\nu_{R}^i$](nu);
\coordinate[below = 1 cm of vaux,label=$t_L$](t);
\coordinate[above right = 1 cm of vaux](v4);
\coordinate[above left = 1 cm of vaux](v3);
\semiloop[vector]{v4}{v3}{180};
\draw[fill=black](v1) circle (.05cm);
\draw[fill=black](v2) circle (.05cm);
\draw[fill=black](v3) circle (.05cm);
\draw[fill=black](v4) circle (.05cm);
\loopv[fermionr]{v1}{v2}{0};
\draw[scalarnoarrow](h1)--(v1);
\draw[scalarnoarrow](v2)--(h2);
\end{tikzpicture}
\end{gathered} .
\end{equation}
The diagram where the left-handed neutrino $\nu_L^3$ pairs with the top quark gives
\begin{equation}
\left.  \delta m_h^2 \right|^{\rm A}_{X,1} =  3 \times 2 \times g_4^2 \left(\frac{2m_t^2}{v^2}\right) I_{\rm A}~, 
\end{equation}
where the factor 3 comes from color, and the 2 is a combinatorial factor that arises from possible ways of contracting the fields.
The diagram involving right-handed neutrinos $\nu_{Rj}$ leads to the following amplitude
\begin{equation}
\left. \delta m_h^2 \right|^{\rm A}_{X,2} =   3 \times 2 \times g_4^2 \left(\frac{2m_t^2}{v^2}\right) \left(\sum_{i=1}^3 |W_R^{3i}|^2 \right) I_{\rm A}~.
\end{equation}
 Similarly, the Higgs couples to the neutrinos, as the following diagrams display
 \begin{equation}\label{diag:tomato}
 \begin{gathered}
\begin{tikzpicture}[line width=1.5 pt,node distance=1 cm and 1 cm]
\coordinate[label=left:$h$](h1);
\coordinate[right = 0.25 cm of h1](v1);
\coordinate[right = 2 cm of v1](v2);
\coordinate[right = 0.25 cm of v2,label=right:$h\text{,}$](h2);
\coordinate[right = 1 cm of v1](vaux);
\coordinate[above left = 0.3 cm of v1,label=above:$\nu_L^3$](tRL);
\coordinate[above right = 0.3 cm of v2,label=above:$\nu_L^3$](tRR);
\coordinate[above =0.1 cm of vaux,label=$X$](Zp);
\coordinate[above = 1 cm of vaux,label=$t_L$](nu);
\coordinate[below = 1 cm of vaux,label=$\nu_R^i$](t);
\coordinate[above right = 1 cm of vaux](v4);
\coordinate[above left = 1 cm of vaux](v3);
\semiloop[vector]{v4}{v3}{180};
\draw[fill=black](v1) circle (.05cm);
\draw[fill=black](v2) circle (.05cm);
\draw[fill=black](v3) circle (.05cm);
\draw[fill=black](v4) circle (.05cm);
\loopv[fermionr]{v1}{v2}{0};
\draw[scalarnoarrow](h1)--(v1);
\draw[scalarnoarrow](v2)--(h2);
\end{tikzpicture}\end{gathered} \
\begin{gathered}
\begin{tikzpicture}[line width=1.5 pt,node distance=1 cm and 1 cm]
\coordinate[label=left:$h$](h1);
\coordinate[right = 0.25 cm of h1](v1);
\coordinate[right = 2 cm of v1](v2);
\coordinate[right = 0.25 cm of v2,label=right:$h$](h2);
\coordinate[right = 1 cm of v1](vaux);
\coordinate[above left = 0.3 cm of v1,label=above:$\nu_R^k$](tRL);
\coordinate[above right = 0.3 cm of v2,label=above:$\nu_R^k$](tRR);
\coordinate[above =0.1 cm of vaux,label=$X$](Zp);
\coordinate[above = 1 cm of vaux,label=$u_R^j$](nu);
\coordinate[below = 1 cm of vaux,label=$\nu_L^3$](t);
\coordinate[above right = 1 cm of vaux](v4);
\coordinate[above left = 1 cm of vaux](v3);
\semiloop[vector]{v4}{v3}{180};
\draw[fill=black](v1) circle (.05cm);
\draw[fill=black](v2) circle (.05cm);
\draw[fill=black](v3) circle (.05cm);
\draw[fill=black](v4) circle (.05cm);
\loopv[fermionr]{v1}{v2}{0};
\draw[scalarnoarrow](h1)--(v1);
\draw[scalarnoarrow](v2)--(h2);
\end{tikzpicture}
\end{gathered} ~.
\end{equation}
The diagram where the $\nu_L^3$ couples with the vector leptoquark gives
\begin{equation}
    \left. \delta m_h^2 \right|_{X,3}^{\rm A} =  3 \times 2 \times g_4^2 \left(\frac{2m_t^2}{v^2} \right)  \left(\sum_{i=1}^3 |W_R^{3i}|^2 \right) I_{\rm A}~, 
\end{equation}
where we sum over the right-handed neutrinos ($i=1,2,3$). The diagram where the right-handed neutrinos couple to the vector leptoquark also gives
\begin{eqnarray}
\left.  \delta m_{h}^2\right|_{X,4}^{\rm A} &=& 3 \times 2 \times g_4^2 \left(\frac{2m_t^2}{v^2}\right) \\
&& \!\!\! \!\!\! \!\!\!  \!\!\!  \times \left(\sum_{i,j,k=1}^3 (W_R^*)^{ij}W_R^{ik} (W_R^*)^{3k}W_R^{3j} \right) I_{\rm A} ~, \nonumber
\end{eqnarray}
where we sum over right-handed neutrinos and right-handed up-type quarks.
Adding all contributions of topology A, {\it i.e.} Diag.~\eqref{diag:potato} and~\eqref{diag:tomato}, 
\begin{eqnarray}
   \left.  \delta m_h^2\right|^{\rm A}_X &=& \left. \delta m_h^2 \right|^{\rm A}_{X,1} +  \left. \delta m_h^2 \right|^{\rm A}_{X,2} +  \left. \delta m_h^2 \right|^{\rm A}_{X,3} +  \left. \delta m_h^2 \right|^{\rm A}_{X,4} \nonumber \\
    &=& 24 \, g_4^2 \left(\frac{2m_t^2}{v^2}\right) I_{\rm A}~.
\end{eqnarray}
One also expects the following diagrams, 
\begin{equation}\label{diag:watermelon}
\begin{gathered}
\begin{tikzpicture}[line width=1.5 pt,node distance=1 cm and 1 cm]
\coordinate[label=left:$h$](h1);
\coordinate[right = 0.25 cm of h1](v1);
\coordinate[right = 2 cm of v1](v2);
\coordinate[right = 0.25 cm of v2,label=right:$h\text{,}$](h2);
\coordinate[right = 1 cm of v1,label=right:$X$](vaux);
\coordinate[above left = 0.3 cm of v1,label=above:$\nu_L^3$](tRL);
\coordinate[above right = 0.3 cm of v2,label=above:$t_L$](tRR);
\coordinate[below left = 0.3 cm of v1,label=below:$\nu_R^i$](bRL);
\coordinate[below right = 0.3 cm of v2,label=below:$t_R$](bRR);
\coordinate[above = 1 cm of vaux](nu);
\coordinate[below = 1 cm of vaux](t);
\draw[vector](t)--(nu);
\draw[fill=black](v1) circle (.05cm);
\draw[fill=black](v2) circle (.05cm);
\draw[fill=black](t) circle (.05cm);
\draw[fill=black](nu) circle (.05cm);
\loopv[fermionr]{v1}{v2}{0};
\draw[scalarnoarrow](h1)--(v1);
\draw[scalarnoarrow](v2)--(h2);
\end{tikzpicture}
\end{gathered} \ 
\begin{gathered}
\begin{tikzpicture}[line width=1.5 pt,node distance=1 cm and 1 cm]
\coordinate[label=left:$h$](h1);
\coordinate[right = 0.25 cm of h1](v1);
\coordinate[right = 2 cm of v1](v2);
\coordinate[right = 0.25 cm of v2,label=right:$h$](h2);
\coordinate[right = 1 cm of v1,label=right:$X$](vaux);
\coordinate[above left = 0.3 cm of v1,label=above:$\nu_R^i$](tRL);
\coordinate[above right = 0.3 cm of v2,label=above:$t_R$](tRR);
\coordinate[below left = 0.3 cm of v1,label=below:$\nu_L^3$](bRL);
\coordinate[below right = 0.3 cm of v2,label=below:$t_L$](bRR);
\coordinate[above = 1 cm of vaux](nu);
\coordinate[below = 1 cm of vaux](t);
\draw[vector](t)--(nu);
\draw[fill=black](v1) circle (.05cm);
\draw[fill=black](v2) circle (.05cm);
\draw[fill=black](t) circle (.05cm);
\draw[fill=black](nu) circle (.05cm);
\loopv[fermionr]{v1}{v2}{0};
\draw[scalarnoarrow](h1)--(v1);
\draw[scalarnoarrow](v2)--(h2);
\end{tikzpicture}
\end{gathered}~,
\end{equation}
which give
\begin{eqnarray}
\left. \delta m_h^2 \right|^{\rm B}_X =   
2 \times  3 \times 2 \times g_4^2  \left(\frac{2m_t^2}{v^2}\right) \left( \sum_{i=1}^3 |W_R^{3i}|^2 \right) I_{\rm B}~, \ \ 
\end{eqnarray}
where, as in the previous cases, the factor $3$ comes from color, a factor $2$ comes from field contractions, and a factor $2$ comes from the two contributions in Diag.~\eqref{diag:watermelon}.

The overall contribution involving the $X$ vector leptoquarks from both topologies is then proportional to 
\begin{equation}
    2 I_{\rm A} + I_{\rm B} = \frac{1}{32\pi^4} m_X^2 \log \left( \frac{m_X^2}{m_h^2}\right)+\cdots\,.
\end{equation}
As expected by general arguments and explicitly shown above, the leading scheme-independent piece scales linearly with the logarithm. 

Therefore, we find 
\begin{equation}
\begin{split}
     \left. \delta m_h^2 \right|_{\rm X}  &=  \left. \delta m_h^2 \right|^{\rm A}_X +  \left. \delta m_h^2 \right|^{\rm B}_X \\
     &= i \frac{3\alpha_s}{\pi^3}  \frac{m_t^2}{v^2} m_X^2 \log \left(\frac{m_X^2}{m_h^2}\right)~.
\end{split}
\end{equation}
Hence, 
\begin{equation}
\left | \frac{\delta m_h^2}{m_h^2} \right | = 5 \times 10^4 \left(\frac{m_X}{100 \text{ TeV}}\right)^2  \left(\frac{\log (m_X^2 / m_h^2)}{13.4}\right)~.
\end{equation}

Analogously to the vector leptoquark case, it can be shown that the leading contribution (proportional to $g_3^2$) coming from the same topologies involving the $Z'$ scales linearly with the logarithm of the heavy neutral gauge boson mass. However, this is slightly smaller than the $X$-boson radiative correction. The heavier mass of the $Z'$ boson is indeed compensated by the prefactors from the $SU(4)$ generators (see Eq.\,\eqref{eq:T15}) and the cosine of the quark-lepton unification angle.

%%%%%%%%%%%%%%%%%%%%%%%%%%%%%%%%%%%%%%%%%%%%%%%%%%%%%%
\subsection{Fermion masses}
%%%%%%%%%%%%%%%%%%%%%%%%%%%%%%%%%%%%%%%%%%%%%%%%%%%%%%%
As manifest in Eq.~\eqref{eq:massrel}, quark-lepton unification (with one Higgs) predicts the same mass matrix for the down-type quarks and charged leptons, which we have not addressed yet. 
Within the context of the ${\cal G}_{\rm PS}$ gauge group and demanding quark-lepton unification, there are two ways to correct the mass relation above at the renormalizable level\footnote{We do not consider corrections from Planck suppressed operators as we expect the scale of quark-lepton unification to be considerably lower than $M_{\rm Pl}$ so that their effect should be irrelevant.}: 
\begin{enumerate}[wide, labelwidth=!, labelindent=0pt]
    \item {\bf Adding extra scalars}. The simplest option of this class is to add a single scalar representation in the adjoint of $\SU(4)$,
    \begin{equation}
        \Phi_{15} = (15, 2, 1/2)_{\rm PS} = \begin{pmatrix} \Phi_8 & \Phi_3 \\ \Phi_4 & 0  \end{pmatrix} + \sqrt{2} H_2 T_{15}\,,
    \end{equation}
    which contains a colored doublet Higgs boson $\Phi_8 \sim (8,2,1/2)_{\rm SM}$, two scalar leptoquarks $\Phi_3 \sim (\bar 3,2,-1/6)_{\rm SM}$ and $\Phi_4\sim (3,2,7/6)_{\rm SM}$, and a \mbox{second} Higgs boson $H_2 \sim (1,2,1/2)_{\rm SM}$. The quantum numbers of $\Phi_{15}$ allow Yukawa-type couplings with the standard model fermions
    \begin{equation}
    \begin{split}
       - {\cal L} \supset & \, Y_1 \bar F_{QL} \widetilde H_1  F_u + Y_2 \bar F_{QL} \widetilde \Phi_{15}  F_u \\
       &+ Y_3 \bar F_{QL} H_1  F_d + Y_4 \bar F_{QL} \Phi_{15} F_d + \text{h.c.}\,,
    \end{split}
    \end{equation}
    where $\widetilde \Phi_{15} = i \sigma_2 \Phi_{15}^*$. 
    Because $\Phi_{15}$ is in the adjoint of $\SU(4)$, unlike the Higgs boson $H_1$, it distinguishes quarks and leptons and allows for a splitting in their masses, such that
    \begin{eqnarray}
        M_u &=& Y_1 \frac{v_1}{\sqrt{2}} + \frac{1}{2\sqrt{3}}Y_2 \frac{v_2}{\sqrt{2}}\,,\\
        M_\nu^D &=& Y_1 \frac{v_1}{\sqrt{2}}-\frac{\sqrt{3}}{2}Y_2 \frac{v_2}{\sqrt{2}}\,,\\
        M_d&=& Y_3 \frac{v_1}{\sqrt{2}} + \frac{1}{2\sqrt{3}}Y_4 \frac{v_2}{\sqrt{2}}\,,\\
        M_e &=& Y_3 \frac{v_1}{\sqrt{2}} - \frac{\sqrt{3}}{2}Y_4 \frac{v_2}{\sqrt{2}}\,.
    \end{eqnarray}
    The vevs are defined as $\langle H^a_1 \rangle = \delta^{a2} v_1  / \sqrt{2}$ and $\langle H_2^a \rangle = \delta^{a2} v_2 / \sqrt{2}$, where $a=1,2$ is a $\SU(2)$ index. The scalar potential that only contains the Higgs doublets is given by
    \begin{equation}
    \begin{split}
  \qquad  V &=  m_{11}^2 H^\dagger_1 H_1 + m_{22}^2 H_2^\dagger H_2 \\
  & \quad - m_{12}^2 (H^\dagger_1 H_2 + H_2^\dagger H_1) + {\cal O}(\lambda)\,,
  \end{split}
    \end{equation}
    where ${\cal O}(\lambda)$ terms arise from quartic interactions that we neglect for simplicity.
    The masses of the standard model Higgs, $m_h$, and the heavier Higgs, $m_H$, after applying the minimization conditions of the potential, are given by
    \begin{equation}
      \qquad   m_h^2 = {\cal O}(\lambda v^2)\,, \quad \text{ and }\quad m_H^2 = \frac{m_{12}^2 v^2}{v_1 v_2}\,,
        \end{equation}
    where $v=\sqrt{v_1^2+v_2^2} = 246$ GeV is the electroweak scale. 
    Taking the determinant of the mass-squared matrix of the Higgs boson, 
    \begin{equation}\label{eq:det}
        \qquad (m_{11}^2 m_{22}^2 - m_{12}^4)  \simeq m_h^2 m_H^2\,.
    \end{equation}
The hierarchy $m_H \gg m_h$ can be achieved when $m_{22}^2 \gg m_{12}^2 \gg m_{11}^2$, which leads to $v_2 \ll v_1$. Here we neglect the quartic terms in the scalar potential for simplicity.
In the discussion below we will show that the radiative corrections to $m_{22}$ are proportional to the mass of the heavy gauge boson.

The complete scalar potential for quark-lepton unification with $\chi$, $H_1$ and $\Phi_{15}$ can be found in Refs.~\cite{FileviezPerez:2013zmv,Faber:2018qon,Faber:2018afz,FileviezPerez:2022fni}. It is straightforward to see that assuming small couplings for the terms enhanced by the vev of $\chi$ is enough to avoid fine-tuning problems. For example
\begin{equation}
  V \supset  \lambda H^\dagger_1 H_1 \chi^\dagger \chi~, 
\end{equation}
does not require any cancellation if $\lambda < \alpha_s \, \pi m_h^2/m_X^2 \simeq 6 \times 10^{-7}(100 \text{ TeV}/m_X)^2$, which is of similar order as the electron Yukawa coupling.

Unlike in the case where there is only a Higgs doublet, singlet under $\SU(4)$, here the vector leptoquarks contribute at one-loop to the Higgs mass through the mixing between the standard model Higgs with $H_2$. Despite $H_2$ being a color singlet, it is embedded in a representation that carries $\SU(4)$ charge. The covariant derivative of an adjoint representation involves the commutator of the $\SU(4)$ vector bosons and the representation itself. The kinetic term, $\text{Tr}\{ (D_\mu \Phi)^\dagger D^\mu \Phi \}$, predicts the following interaction
    \begin{equation}\label{eq:58}
    {\cal L} \supset \frac{2}{3} g_4^2 X_\mu^\dagger X^\mu H_2^\dagger H_2~.
    \end{equation}
 Using this term in the Lagrangian, the one-loop contribution of the vector leptoquarks to the Higgs mass is suppressed by the mixing angle between the two Higgs bosons 
  \vspace{-0.2cm}
    \begin{equation}
\begin{split}
 &  \begin{gathered} 
\begin{tikzpicture}[line width=1.5 pt,node distance=1 cm and 1 cm]
\coordinate[label=left:$h$](H1);
\coordinate[right = 1.5 cm of H1](v1);
\coordinate[right= 1.5 cm of v1,label=right:$h$](H2);
\coordinate[above=0.4 cm of v1,label=$\, \, \,X$](Zp);
\draw[fill=black](v1) circle (.05cm);
\draw[decorate, decoration={snake}] 
        (v1) arc (-90:270:0.7);
\draw[scalarnoarrow](H1)--(v1);
\draw[scalarnoarrow](v1)--(H2);
\end{tikzpicture}
\end{gathered} \\
 \delta {m_h^2} 
&=-  \frac{\alpha_s}{3\pi} \sin^2 \theta \ m_{X}^2 \log \left(\frac{m_{X}^2}{m_h^2}\right).
\label{eq.59}
\end{split}
\end{equation}
The mixing angle, $\theta$, is defined by
$h = \cos \theta \ {\rm Re} (H_1^2) + \sin \theta \ {\rm Re} (H_2^2)$.
When $v_2 \ll v_1$, $\theta$ and thus $\delta m_h^2$ are suppressed.  However, $v_2$ cannot be arbitrarily small since it plays the role of correcting the $M_e = M_d$ relation. For perturbative Yukawa couplings, $v_2 \gtrsim {\cal O}(\text{GeV})$ is required to reproduce the observed masses for the bottom quark and tau lepton. Such a hierarchy allows a $H$ as heavy as $m_{H} v \sim m_{11}m_{22}(v_1/v_2)$ without \mbox{requiring} any tuning. It also allows to relax the one-loop correction to the Higgs mass in Eq.~\eqref{eq.59} to 
\begin{equation}
    \left | \frac{\delta m_h^2}{m_h^2}\right| = 20 \left(\frac{m_{X}}{100 \text{ TeV}}\right)^2 \left(\frac{v_2/v_1}{10^{-2}}\right)^2 \!\!\! \left(\frac{\log (m_X^2 / m_h^2)}{13.4}\right)~.
\end{equation}
Notice that Eq.~\eqref{eq:58} implies a large radiative correction to $m_{22}$ consistent with the hierarchy $m_{22}^2 \gg m_{12}^2 \gg m_{11}^2$ required for $v_2 \ll v_1$.

In this theory, both $Y_3$ and $Y_4$ are needed to correct the charged leptons and down-type quark masses and allow for a light (100 TeV) $X$ vector leptoquark. In the limit where $v_2 \ll  v_1$, the relation  $M_u=M_\nu^D$ is preserved. We expect that the amount of tuning required in this limit is still given by the two-loop processes discussed earlier.

\item {\bf Adding extra fermions}. Another possibility consists of adding vector-like fermions with quantum numbers such that their mixing with the standard model fermions is allowed. Thus, the mass matrix relation $M_d = M_e$ can be corrected to match the observed values of the charged leptons and down-type quarks. 
Notice that the above relation can be corrected, as shown for example in Ref.~\cite{Dolan:2020doe}, without modifying the quark-lepton unification prediction $M_u = M_\nu^D$. Additionally, the lower bound on the vector leptoquark mass $m_X$, and consequently $m_{Z'}$ (see Eq.~\eqref{eq:MZMXratio}), can be considerably relaxed, as the unitarity condition on the mixing matrices is lifted. See Ref.~\cite{Dolan:2020doe} for a recent study on this possibility. The bound on the vector leptoquark mass in this case can be as low as a few TeV, which would also relax the tuning from the two-loop contributions
% (see Eq.~\eqref{eq:2loopQL}).%to 
\begin{equation}\label{eq:newtuning}
\left | \frac{\delta m_h^2}{m_h^2} \right | = 1 \times 10^2 \left(\frac{m_X}{6 \text{ TeV}}\right)^2  \left(\frac{\log (m_X^2 / m_h^2)}{7.7}\right)~.
\end{equation}
\end{enumerate}
%
%%%%%%%%%%%%%%%%%%%%%%%%%%%%%%%%%%%%%
\section{CONCLUDING REMARKS}
The hierarchy problem, arising from the quadratic dependence of the Higgs mass parameter on the momentum cut-off, has motivated many proposed extensions of the SM. However, the absence of new particle discoveries at the LHC has raised doubts about the relevance of this issue. While quadratic divergences do not appear in all regulator approaches, certain extensions of the SM exhibit a quadratic dependence of the Higgs mass on the other masses of the theory, independent of the regularization scheme. This motivates the principle of finite naturalness, which demands that these quantities are not too much larger than the Higgs mass.

In this article, we explored whether Pati-Salam models with $\SU(3)_C$ embedded in $\SU(4)$ respect the principle of finite naturalness. We showed that the minimal realization of quark-lepton unification leads to radiative contributions to the Higgs boson mass squared $\delta m_h^2 / m_h^2 \gtrsim {\cal O}(10^4)$, which is problematic regarding finite naturalness. We computed the two-loop corrections to the Higgs mass involving the vector leptoquark and showed that the naively leading contributions from these corrections cancel in the minimal theory for quark-lepton unification. Something analogous happens for the expected dominant contributions from the two-loop corrections involving the new neutral gauge boson, which cancel out. We also briefly mentioned extensions of the minimal model containing extra fermions where the level of tuning can be relaxed to $\delta m_h^2 / m_h^2 \sim {\cal O}(10^2)$ by lowering the bound on the heavy gauge boson masses.

{\small{\textit{\bf{Acknowledgments:}}}
We thank Federico Cima, Andreas Helset, Ryan Plestid, and Michele Tarquini for useful discussions.
The work of P.F.P. is supported by the U.S. Department of Energy, Office of Science, Office of High Energy Physics, under Award Number DE-SC0024160.
The work of C.M., S.P., A.T., and M.B.W. is supported by the U.S. Department of Energy, Office of Science, Office of High Energy Physics, under Award Number DE-SC0011632, and by the Walter Burke Institute for Theoretical Physics. M.B.W. thanks Perimeter Institute for their hospitality during the later stages of this work.}

%%%%%%%%%%%%%%%%%%%%%%%%%%%%%%%%%%%%%

\bibliography{FN}

\end{document}